\documentclass[11pt, conference]{IEEEtran}
\IEEEoverridecommandlockouts
\usepackage{cite}
\usepackage{amsmath,amssymb,amsfonts}
\usepackage{algorithmic}
\usepackage{graphicx}
\usepackage{textcomp}
\usepackage{xcolor}
\usepackage{hyperref}
\usepackage{float}
\usepackage{url}
\usepackage{comment}

\begin{document}

\title{Performance of Confidential Computing GPUs\\

\thanks{This study was supported by grant RYC2023-043553-I, funded by MICIU/AEI/10.13039/501100011033 and ESF+ as well as the HORIZON-MSCA-2021-SE-01-01 project Cloudstars (g.a. 101086248). The publication is also part of NEREIDAS TSI-100120-2024-13 funded by EU NextGenerationEU (PRTR).}
}

\author{\IEEEauthorblockN{Antonio Martínez Ibarra\IEEEauthorrefmark{1}, 
Julian James Stephen\IEEEauthorrefmark{2},
Aurora González Vidal\IEEEauthorrefmark{1}, \\
K. R. Jayaram\IEEEauthorrefmark{2},
Antonio Fernando Skarmeta Gómez\IEEEauthorrefmark{1}}
\IEEEauthorblockA{\IEEEauthorrefmark{1}University of Murcia,
\IEEEauthorrefmark{2}IBM Research}}

\maketitle

% IEEE copyright statement - added to submit to arXiv
\vspace{-1.5em}
\noindent
\footnotesize
© 2025 IEEE. Personal use of this material is permitted. Permission from IEEE must be obtained for all other uses, in any current or future media, including reprinting/republishing this material for advertising or promotional purposes, creating new collective works, for resale or redistribution to servers or lists, or reuse of any copyrighted component of this work in other works.
\vspace{0.5em}

\begin{abstract}
This work examines latency, throughput, and other metrics when performing inference on confidential GPUs. We explore different traffic patterns and scheduling strategies using a single Virtual Machine with one NVIDIA H100 GPU, to perform relaxed batch inferences on multiple Large Language Models (LLMs), operating under the constraint of swapping models in and out of memory, which necessitates efficient control. 
The experiments simulate diverse real-world scenarios by varying parameters such as traffic load, traffic distribution patterns, scheduling strategies, and Service Level Agreement (SLA) requirements. The findings provide insights into the differences between confidential and non-confidential settings when performing inference in scenarios requiring active model swapping.

Results indicate that in No-CC mode, relaxed batch inference with model swapping latency is 20–30\% lower than in confidential mode. Additionally, SLA attainment is 15–20\% higher in No-CC settings. Throughput in No-CC scenarios surpasses that of confidential mode by 45–70\%, and GPU utilization is approximately 50\% higher in No-CC environments. Overall, performance in the confidential setting is inferior to that in the No-CC scenario, primarily due to the additional encryption and decryption overhead required for loading models onto the GPU in confidential environments.
\end{abstract}


\section{Introduction}

Confidential computing addresses the need to protect data in use by performing computations within a hardware-based, attested Trusted Execution Environment (TEE) \cite{Confidential_Computing_2022a}. These secure and isolated environments prevent unauthorized access or modification, ensuring data integrity, data confidentiality, and code integrity \cite{Confidential_Computing_2022b}. Confidential Virtual Machines (CVMs), a key implementation of confidential computing, operate within such TEEs to shield their code and data from hypervisors, host operating systems, and other components of the TEE's host environment \cite{Confidential_Computing_2022c}.

The proliferation of artificial intelligence (AI) has increased the necessity of protecting not only data but also the models themselves, including their architectures, weights, activations, and outputs \cite{Linux_Foundation_2024}. Secure AI inference is particularly challenging because models process sensitive data while requiring substantial computational resources \cite{Naveed_Khan_Qiu_Saqib_Anwar_Usman_Akhtar_Barnes_Mian_2024}. The additional overhead of running AI inference within confidential environments—due to encryption, decryption, and attestation mechanisms—further complicates performance optimization. Efficient workload scheduling in CVMs is crucial for balancing security and performance while meeting Service Level Agreements (SLAs).

While confidential computing on CPUs has been extensively studied, research on the performance of confidential GPUs and GPU-based TEEs for AI inference remains relatively scarce. GPUs play a critical role in accelerating AI workloads, particularly in deep learning, making their secure deployment a pressing concern. Unlike CPUs, GPUs introduce additional complexities, including memory access patterns, multi-stream parallelism, and specialized encryption mechanisms, which can significantly impact performance in confidential computing environments. Understanding the trade-offs between security and performance in confidential GPU inference is essential for optimizing AI deployments in production environments.

To address this gap, this study investigates the performance implications of executing AI inference workloads in Confidential Compute (CC) mode versus Non-Confidential (No-CC) mode on GPUs. Experiments will evaluate key trade-offs, including latency, throughput, SLA attainment, and GPU utilization, using realistic AI workloads. While Large Language Models (LLMs) serve as a representative case study, the insights gained extend to a broader range of deep learning applications in fields such as computer vision, scientific computing, and healthcare. This analysis lays the groundwork for improving scheduling strategies and provisioning AI workloads efficiently in confidential GPU environments.

\section{Background and Related Work}

\subsection{Types of Inference}

Although many machine learning (ML) engineers try to shorten inference time through
techniques like quantization, specialized model compilation, etc., in practice,
inference can be classified into different types based on response latency requirements:

\begin{itemize}
    \item \textbf{Hard Real-time Inference:} In these applications, failure to perform inference
    within the required timeframe will likely either be life-threatening or cause a substantial monetary loss. This type of inference involves dedicated hardware which is not shared and is consequently not
    ``interesting'' from a performance optimization and scheduling point of view. It is used in latency-critical applications such as autonomous vehicles, high-frequency trading, and satellite systems.
    
    \item \textbf{Soft Real-time Inference:} Also demands fast response times but can tolerate some level of latency without compromising overall system reliability or safety. This type of inference is widely used in online services like chatbots and applications such as voice assistants and recommendation systems. Not meeting SLAs here can lead to customer dissatisfaction but seldom loss of life or large monetary damages.
     
    \item \textbf{Relaxed Inference:} Allows for more flexible inference times, ranging from minutes to hours.
    Relaxed inference is widely used, either (i) when the objective is to keep processing costs low by using spot cloud instances, examples include invoice processing, understanding receipts and processing expense reports (for which there is usually 24+ hours, if not a few days), or (ii) in cases where the same data
    item is fed to multiple models in sequence, e.g., in medical diagnosis or weather prediction.
    
    \item \textbf{Best-effort Inference:} Occurs in scenarios where no dedicated resources are allocated for inference. This is commonly found in peer-to-peer systems and community-driven environments, where inference is performed over a wide-area network with variable availability. This type of inference does not have SLAs. % and     is also  
\end{itemize}

\subsection{Trusted Execution Environments (TEEs)}

\begin{figure}[htbp]
    \centering
    \includegraphics[width=.40\textwidth, keepaspectratio=true]{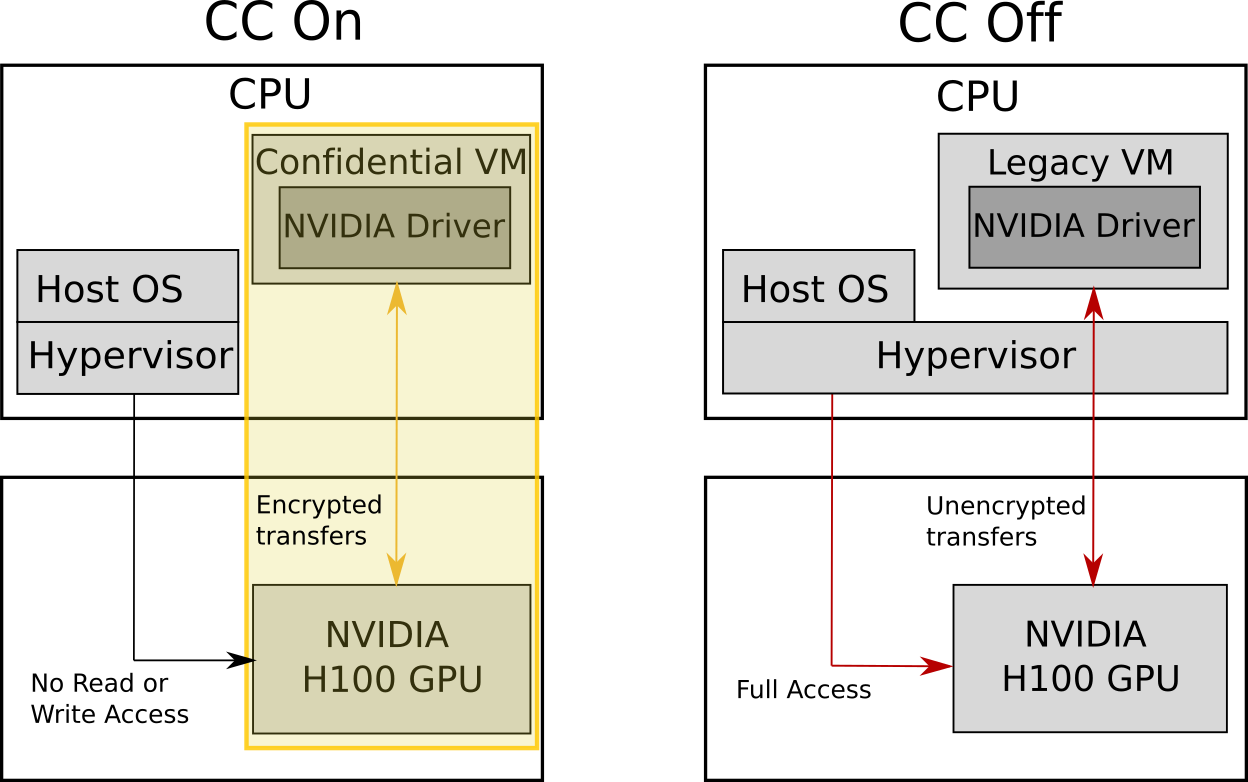}
    \caption{Confidential Computing with NVIDIA H100 (modified from \cite{NVIDIA_H100_2023})}
    \label{fig:diagram_CC}
\end{figure}

TEEs provide secure execution environments with dedicated memory and storage protections, enabling computations on sensitive data while preventing unauthorized access. CPU-based TEEs, such as Intel TDX and AMD SEV, utilize hardware extensions to manage and encrypt memory, ensuring data protection during execution~\cite{IntelTdx}. Confidential GPUs, such as the NVIDIA H100~\cite{NVIDIA_H100_2023}, extend CPU-based TEEs by enabling secure data transfer and control signal communication between the CPU and GPU. This key difference is shown in Fig.~\ref{fig:diagram_CC}. Without CC mode, data transfers between the CPU and GPU can be leaked by an un-trusted hypervisor.

The H100 GPU relies on \textbf{secure boot}~\cite{secureboot} and \textbf{hardware attestation}~\cite{remoteattestation,attestation2}. 
\textbf{Secure boot} ensures that only verified firmware and software are permitted to initialize the GPU, mitigating risks from malicious code injections. \textbf{Hardware attestation} provides cryptographic proof of the GPU’s secure state, allowing external systems to verify the integrity of the execution environment~\cite{NVIDIA_H100_2023, NVIDIA_Hopper_2023}. To enable the CC capabilities of the NVIDIA H100 GPU, specific host side requirements must also be met:
(i) \textbf{CPU TEE:} The host system must support CPU TEEs such as AMD SEV-SNP or Intel TDX to securely manage CVMs, and (ii) \textbf{Secure Hypervisor:} A hypervisor with CC support (e.g., KVM) to facilitate CPU, memory and other resource allocations securely. These components work in tandem with the H100’s Hopper architecture to provide end-to-end security, making confidential GPU computing viable for privacy-critical  multi-tenant cloud environments. 

\subsection{Related Work}

Recent research in CC has explored the performance overheads of extending TEEs to GPU-accelerated workloads. One study \cite{Mohan_Ye_Franke_Srivatsa_Liu_Gonzalez_2024} evaluated Intel TDX with NVIDIA H100 GPUs on real-world applications, including LLMs, finding that while CC incurs penalties—especially for smaller models and batch sizes—pipelining can match No-CC performance for larger models. Another study \cite{Mo_Tarkhani_Haddadi_2024} examined TEE limitations in ML workloads, suggesting improvements like partitioned execution and TEE-aware ML designs. Still, the authors highlight that balancing privacy guarantees with performance remains a key challenge.

CC in HPC has been studied with findings showing that current TEEs struggle to balance security and performance for demanding workloads~\cite{Akram_Akella_Peisert_Lowe-Power_2022}. A related study~\cite{Zhu_Yin_Deng_Zhou_2024} evaluated secure LLM inference on NVIDIA H100 GPUs, finding low GPU overhead but identifying encrypted data transfers as the primary bottleneck for scalable confidential inference.

Various strategies have been proposed to optimize LLM inference scheduling (LLM serving). \cite{Li_Zheng_Zhong_Liu_Sheng_Jin_Huang_Chen_Zhang_Gonzalez_2023} introduce inter-operator parallelism for automatic parallelization and placement of models for serving, though they assume known arrival distributions of inference requests and omit batching. In contrast, \cite{Wang_Wang_Gao_Zhang_Chen_Ng_Ooi_2018} use greedy batching - processing at the max batch size unless a time constraint is about to be violated. Another study \cite{NSDI17} presents a dynamic batch sizing approach, adjusting batch sizes to optimize performance—growing the batch when latency constraints allow and shrinking it when necessary to prevent excessive delays for early queries in a batch. Further, \cite{Ye_Gao_Hu_Sun_Wang_Luo_Zhang_Wen_2024} emphasize that leveraging batch size effectively is key to optimizing throughput.

Despite recent advances, no prior work has explored scheduling techniques tailored for CC environments or compared them to No-CC settings under identical conditions. Existing research also overlooks the challenges of frequent model loading in confidential settings with relaxed inference workloads. This paper addresses these gaps by analyzing scheduling trade-offs in CC-based GPU inference.

\section{Experimental Setup}

This study follows a structured methodology to evaluate the performance of CC and No-CC environments for GPU-based inference. The experiments involve generating inference requests, profiling model performance, implementing scheduling strategies, and measuring key system metrics.

\subsection{Workflow\label{sec:workflow}}

We consider the scenario where the system under evaluation receives relaxed inference requests against different AI models. The evaluation process consisted of the following steps:

\begin{enumerate}
    \item \textbf{Generating Test Data for Inference Batches.}  
    Sample inference requests were generated using Instructlab \cite{instructlab}, which produces a \texttt{jsonl} file containing a specified number of prompts and their corresponding answers. The \texttt{jsonl} file is then converted into a \texttt{json} document, where each entry includes a text prompt and its designated model for inference.

    \item \textbf{Profiling Model Loading and Unloading Times.}  
    To understand the overhead introduced by model swapping, we measured the time taken to load and unload LLMs of different sizes. The process involved loading model parameters and tokenizers onto the GPU, recording the latency, clearing GPU memory, and repeating the process to obtain consistent measurements.

    \item \textbf{Profiling Inference Batches.} Inference batch size represents the number of requests that are processed in a single pass through a single model. Batching enables us to reuse parts and weights of the model that are already loaded into the GPU’s on chip memory. We test each model with progressively larger batch sizes until the GPU runs out of memory. For each batch, we recorded the model type, batch size, processing time, throughput, and various CPU/GPU parameters to analyze computational efficiency.

    \item \textbf{Implementing and Evaluating Scheduling Strategies.}  
    Multiple scheduling techniques were developed to assess system performance under varying traffic patterns and constraints. These strategies aim to balance latency, throughput, and SLA attainment.

    \item \textbf{Measuring System Performance.}  
    Experiments were conducted to assess throughput, latency, SLA attainment, and GPU utilization under different input traffic patterns and scheduling strategies. The same set of experiments was performed in both CC and No-CC environments.
\end{enumerate}

Each experiment was conducted over a \textbf{20-minute run}, where inference requests were generated following a predefined traffic distribution and input rate. A \textbf{single VM with one GPU} was used, capable of serving one model at a time.

\subsection{Components}

The experiments involved three core components:

\begin{itemize}
    \item \textbf{Request Generation (Python Script):} Simulates incoming inference requests based on predefined traffic patterns and rates while logging all transactions.
    \item \textbf{Inference Handling (Flask API):} Batches incoming requests according to specified scheduling strategies and processes them using the selected LLM.
    \item \textbf{Execution (Bash Script):} Iterates through SLA values, traffic parameters, and scheduling strategies, calling the Python script for each combination. The script stores results in \textbf{CSV files} containing: (i) \textbf{Request-level details:} Arrival time, inference timestamp, model used, batch size, and latency. (ii)\textbf{Throughput metrics:} Batch processing time, inference throughput, and overall throughput. (iii) \textbf{System monitoring logs:} CPU and GPU metrics, model switches, total runtime, and inference time.
\end{itemize}

\subsection{Parameters}
We now describe the parameters studied as part of the experimental evaluation.

\subsubsection{Input Traffic Distributions} 
Request arrival patterns play a crucial role in inference performance and scheduling. We picked the following traffic distributions because of their pervasiveness and importance for inference~\cite{Li_Zheng_Zhong_Liu_Sheng_Jin_Huang_Chen_Zhang_Gonzalez_2023}:

\begin{itemize}
    \item \textbf{Gamma Distribution:} Characterized by irregular inter-arrival times, where some requests occur in rapid succession while others are spaced apart. This distribution is common in human-driven interactions and event-driven systems, such as web browsing, financial trading, or IoT devices sending telemetry data at varying intervals. In \textbf{CC environments}, unpredictable request arrivals can amplify inefficiencies due to model loading delays.
  
    \item \textbf{Bursty Traffic:} This pattern alternates between periods of intense activity and idle phases, mimicking real-world workload surges. Promotional campaigns, time-sensitive applications or sudden popularity of a funny camera filter can cause sudden spikes in traffic to specific AI models. In CC environments, \textbf{frequent model switching and encryption overheads} may worsen performance during bursts.

    \item \textbf{Ramp Traffic:} Gradually increases to a peak before tapering off, representing ramp-up and ramp-down phases. This pattern models AI workloads in scheduled pipelines or system warm-ups. In CC settings, \textbf{increasing workloads may expose performance bottlenecks} in model swapping.
\end{itemize}

\subsubsection{Traffic Pattern Mean} 
\begin{figure}
    \centering
    \includegraphics[width=.45\textwidth]{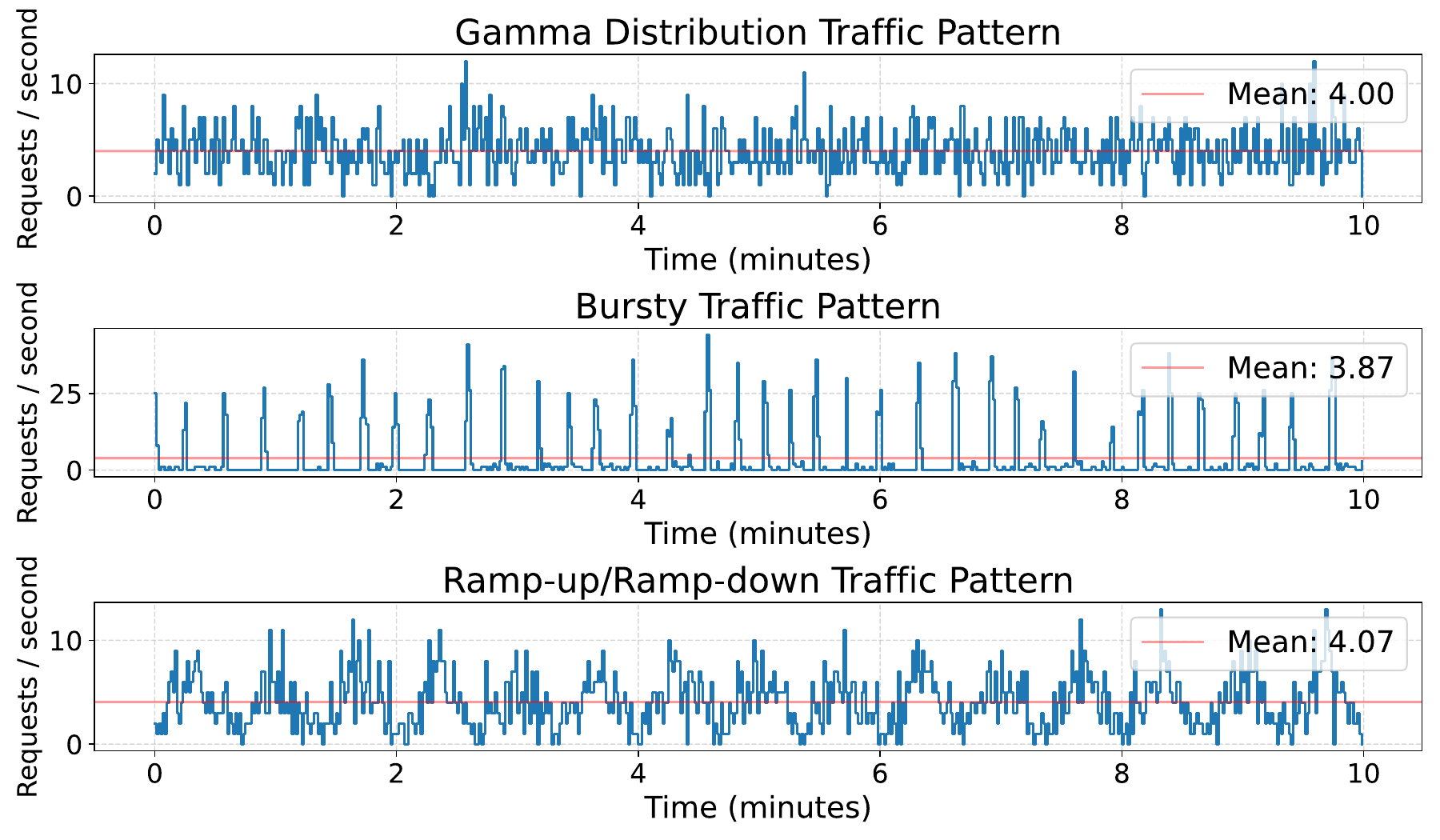}
    \caption{Example of input traffic distributions with a mean of 4 requests per second.}
    \label{fig:input_distributions}
\end{figure}

To ensure a fair comparison, each traffic pattern was configured to generate the same mean requests per second at the end of the full workload. This ensures a consistent request load across distributions. Fig. ~\ref{fig:input_distributions} illustrates the three distributions for a traffic pattern mean of 4.

\subsubsection{Service Level Agreements (SLAs)}

Inference requests must be processed within predefined time constraints, beyond which they are considered unfulfilled. We evaluate the impact of scheduling strategies on SLAs of \textbf{40, 60, and 80 seconds}, which align with relaxed inference workload SLAs. Later, these SLAs will be referred to as SLA 40, SLA 60 or  SLA 80, respectively.

\subsubsection{Scheduling Plans}

Inference requests are queued in order of arrival with one queue for every model. Scheduling plans determine how requests in these queues are batched and sent to the model for processing. We define \textbf{OBS} (Optimal Batch Size) for a model as the batch size that gives maximum throughput for that specific model determined from prior profiling (\autoref{sec:workflow},~\autoref{sec:batchprofiling}).
The following plans were designed, leveraging different batching mechanisms:

\begin{itemize}
    \item \textbf{Best Batch:} Waits until the number of requests in a batch matches the OBS for the corresponding model.
    \item \textbf{Timer:} Introduces a maximum wait time (or timeout) to ensure inference occurs within SLA constraints. If the wait time of a request exceeds the configured maximum wait time, the batch is processed immediately irrespective of the number of requests in the batch.
    \item \textbf{Partial Batch:} Always processes incomplete batches for the currently loaded (in GPU) model before switching to process requests for another model.  
    \item \textbf{Select Batch:} 
    Dynamically selects batch sizes based on prior arrival rates and SLA constraints. Since the time to fill a batch (batch\_accumulation\_time) is the batch size divided by the arrival rate, which is an estimate calculated from past request arrival frequency, we can state:
    \[
    \text{batch\_size} = \text{batch\_accumulation\_time} \times {\text{arrival\_rate}}
    \]
    In order to meet the latency constraint prescribed the the SLAs, the invariant: 
    \[
    \text{batch\_accumulation\_time} <=  {\text{arrival\_rate}}
    \] must hold. Therefore, we can get the maximum batch size that meets the SLA as:
    \[
    \text{batch\_size} \leq \text{arrival\_rate} \times \text{desired\_latency}
    \]
    Here, desired\_latency is determined as the set SLA minus an estimated time for model loading and batch processing.
\end{itemize}

The four scheduling plans above can further be combined in multiple ways to create many scheduling strategies. These scheduling strategies ultimately go to the scheduler and is used to process incoming inference requests. 
The first strategy, ``Best Batch'' is a direct application of the ``Best Batch'' plan, designed with the aim of setting a baseline. The second strategy ``Best Batch + Timer'' tries to create batches of size OBS, but forces the batch to be processed if the wait times exceed configured timeout. The third strategy is ``Select Batch + Timer'' and the idea here is similar to the previous, but with dynamically changing batch sizes. The last one is ``Best Batch + Partial Batch + Timer'' and builds on the second strategy by adding an additional processing step to execute partially filled batches before swapping the model, aiming to increase throughput while minimizing swaps.
\begin{table}
\caption{Scheduling strategies}
\begin{tabular}{|p{0.38\linewidth}|p{0.52\linewidth}|}
\hline
\textbf{Strategy}                  & \textbf{Goal}                                       \\ \hline
Best Batch                         & Set a baseline                                      \\ \hline
Best Batch + Timer                 & Meet SLAs while maintaining a reasonable throughput \\ \hline
Select Batch + Timer               & Meet SLA better                                     \\ \hline
Best Batch + Partial Batch + Timer & Meet SLAs and achieve a higher throughput           \\ \hline
\end{tabular}
\label{tab:scheduling_strategies}
\end{table}
The goals of these strategies are summarized in \autoref{tab:scheduling_strategies}.
The models used for the evaluations are available in Huggingface \cite{HuggingFaceAI2024} and summarized in \autoref{tab:models}.

       \begin{table}[!b]
   \caption{Models used for evaluation}
        \begin{tabular}{|p{0.38\linewidth}|p{0.52\linewidth}|}
        \hline
        \textbf{Model Name}                  & \textbf{Size}                                       \\ \hline
        Llama-3.1-8B & 8B (16.07 GB)  \\ \hline
        gemma-7b  & 7B (17.07 GB)  \\ \hline
        granite-7b-base & 7B (26.98 GB)  \\ \hline
            \end{tabular}
    \label{tab:models}
    \end{table}

\subsection{Model Load Time and Batch Size Profiling}

\subsubsection{Model Load Time Profiling}

\begin{figure}
    \centering
    \includegraphics[width=.445\textwidth]{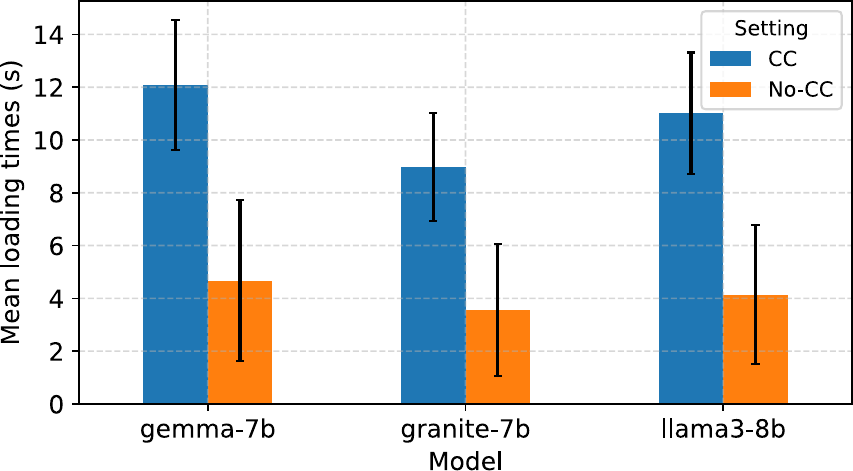}
    \caption{Model loading times in CC and No-CC environments.}
    \label{fig:loading_times}
\end{figure}

Model load time profiling, illustrated in Fig.~\ref{fig:loading_times} measures model \textbf{loading and unloading times}, recorded over multiple iterations. The loading process includes tokenizer and model parameter initialization, as well as GPU memory allocation and I/O overheads.
Note that time spent for code initialization (e.g., of PyTorch framework) is not included in the model loading time.
Results indicate that \textbf{model load time is significantly higher in CC environments}. In our evaluations, model unload times were similar in both scenarios and negligible,ranging between 0.004 and 0.01 seconds.

\subsubsection{Batch Size Profiling\label{sec:batchprofiling}}

\begin{figure}[!b]
    \centering
    \includegraphics[width=.45\textwidth]{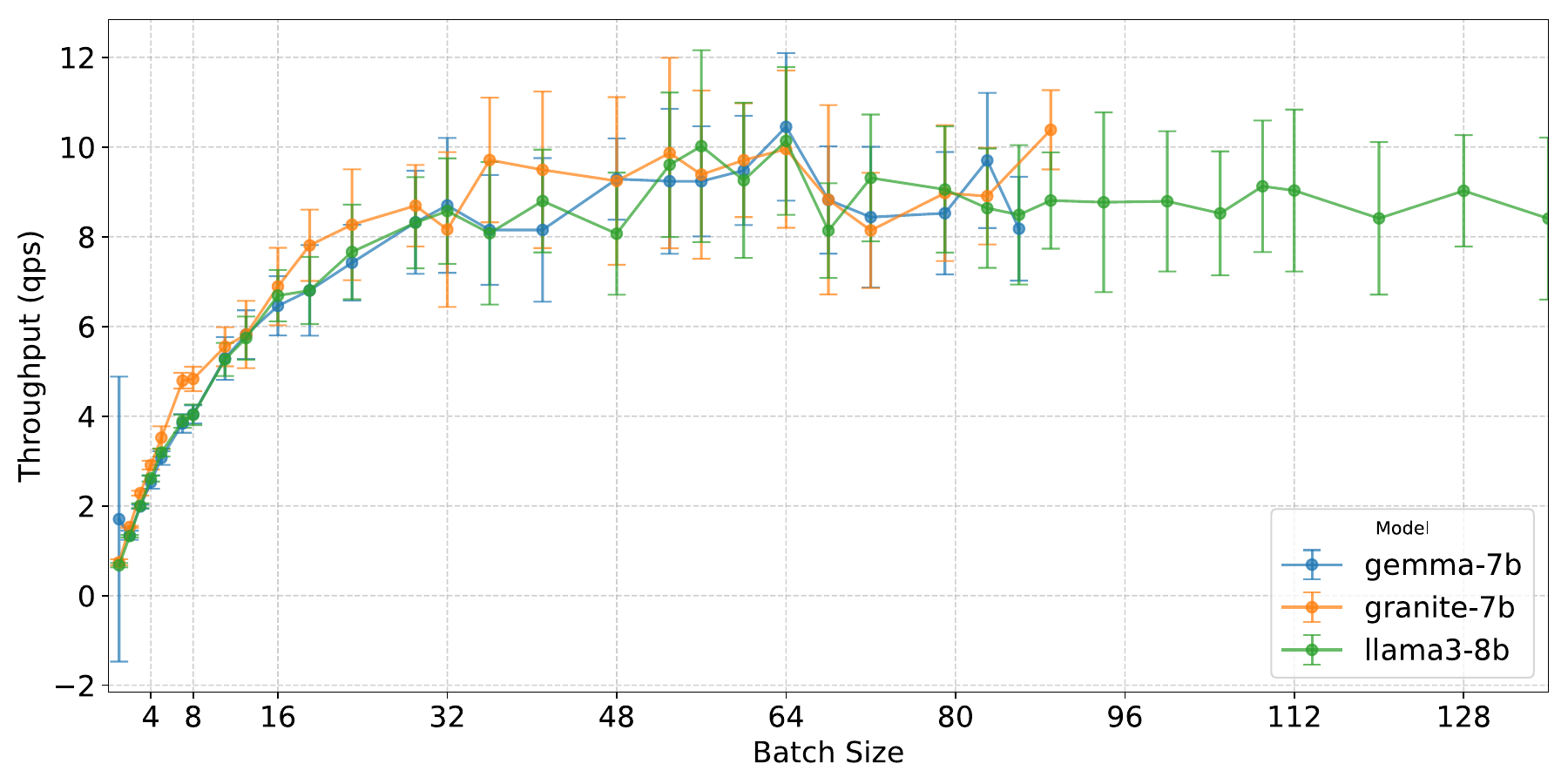}
    \caption{Inference throughput vs. batch size.}
    \label{fig:batch_profiling}
\end{figure}

Batch size profiling determines the OBS for each model by increasing batch sizes until an out-of-memory error occurs. During this process, we recorded model names, batch sizes, processing times, throughput, and CPU/GPU utilization. Result of this profiling are shown in Fig. \ref{fig:batch_profiling}.
For consistency, \textbf{output sequence length is fixed at 50 tokens} across all experiments.

In summary, this experimental setup ensures a controlled and reproducible evaluation of GPU-based TEEs for inference. By systematically profiling models, assessing scheduling strategies, and analyzing workload performance under different conditions, we aim to provide insights into optimizing AI inference in CC environments.
The source code for each stage of the process is publicly available at \url{https://github.com/Antonio-MI/sincere}.

\section{Experiments and Results}
\label{sec:experiments}

The experiments were conducted on a virtual machine equipped with an NVIDIA H100 80GB HBM3 GPU, running driver version 550.54.14, CUDA 12.4, PyTorch 2.4.0+cu121, and Python 3.10.12. The system was based on the Ubuntu 22.04-amd64 Linux distribution.

This section presents a comparative analysis of latency and SLA attainment, throughput, and GPU usage between CC and No-CC settings.

\subsection{Latency and SLA Attainment}

\begin{figure}
    \centering
    \includegraphics[width=.48\textwidth]{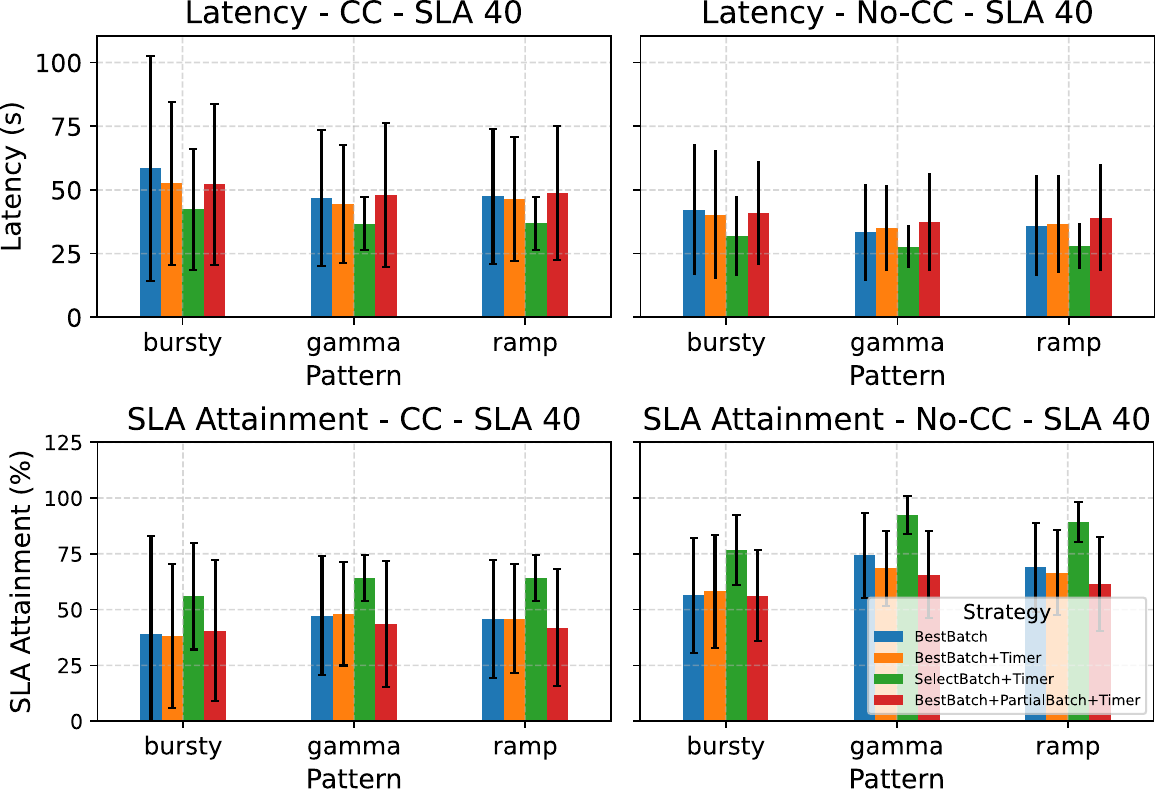}
    \caption{Latency and SLA attainment across different traffic patterns.}
    \label{fig:latency}
\end{figure}

Inference request latency is defined as the time elapsed from when a request is sent by the user until it is dispatched by the server after completing inference. Across all input distributions, inference latency in No-CC mode is consistently \textbf{20–30\% lower} than in CC mode. Among the tested distributions, the \textbf{bursty traffic pattern exhibits the highest latency}, performing worse than both gamma and ramp distributions. This trend remains consistent for higher SLA values, as shown in Fig. \ref{fig:latency}.
 
SLA attainment, measured as the percentage of requests completed within the specified SLA limit, follows an expected pattern. At SLA 40, completion rates are lower due to insufficient time for request processing, and at SLA 60 and 80, completion rates improve significantly. 
Since SLA attainment is inversely correlated with latency, the bursty pattern also records the lowest attainment percentage.

Among the tested scheduling strategies, \textit{``SelectBatch+Timer'' achieves the best performance}. This is because, unlike other methods that prioritize larger batch sizes (leading to longer wait times), this strategy processes smaller batches more frequently, reducing overall latency.

Another observed difference between CC and No-CC settings is the number of \textit{model swaps}. The swap count is slightly higher in No-CC mode, implying that the system can handle more model loads and unloads within the same experimental duration.

The completion rates for CC vs. No-CC settings at different SLA values are:
\begin{itemize}
    \item SLA 40: 50\% (CC) vs. 70\% (No-CC)
    \item SLA 60: 70\% (CC) vs. 85\% (No-CC)
    \item SLA 80: Above 90\% for both CC and No-CC
\end{itemize}

\subsection{Throughput}

\begin{figure}[htbp]
    \centering
    \includegraphics[width=.48\textwidth]{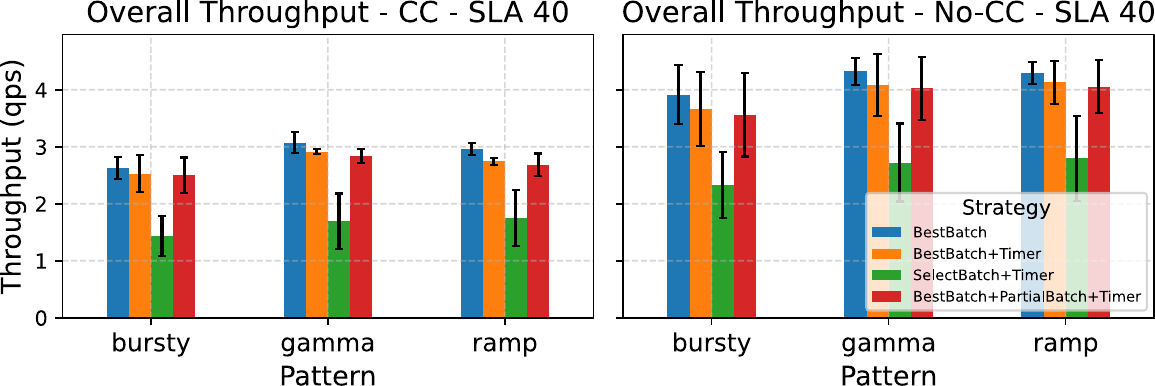}
    \caption{Throughput comparison between CC and No-CC settings.}
    \label{fig:throughput}
\end{figure}

Throughput is measured as the total number of requests processed divided by the total runtime. The results for SLA 40 are shown in Fig. \ref{fig:throughput}.

Among the tested scheduling strategies, the three incorporating ``BestBatch'' logic achieve similar throughput, significantly outperforming ``SelectBatch.'' Interestingly, ``BestBatch+PartialBatch+Timer'' did not yield the highest throughput because it is impractical to configure all scheduling parameters for it optimally.

Across different input patterns, throughput remains relatively consistent, although \textbf{bursty traffic exhibits slightly lower performance}. Comparing CC and No-CC settings, throughput in No-CC mode is \textit{45–70\% higher} than in CC mode. However, the processing rate during inference—defined as the number of requests processed within the time elapsed from model processing start to completion—remains consistent across all input rates, patterns, and scheduling strategies. This suggests that the primary bottleneck in CC mode is not inference execution but rather the overhead associated with model swapping.

\subsection{GPU Utilization}

\begin{figure}[htbp]
    \centering
    \includegraphics[width=.48\textwidth]{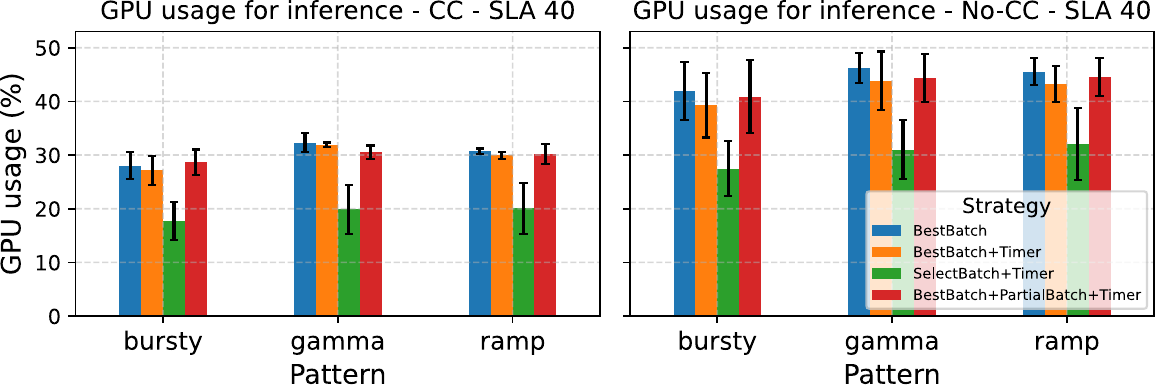}
    \caption{GPU utilization comparison between CC and No-CC settings.}
    \label{fig:gpu_usage}
\end{figure}

GPU utilization is measured as the percentage of total runtime during which the GPU actively performs inference. The objective is to maximize GPU usage for inference workloads. The results (Fig. \ref{fig:gpu_usage}) indicate that GPU utilization in No-CC mode is $\approx$\textit{50\% higher} than in CC mode. However, in both cases, utilization remains \textit{below 50\%} (Fig. \ref{fig:gpu_usage}).

A key question arises: \textit{Where is the remaining time spent?}  
Analysis reveals that most of the unused time is dedicated to \textit{loading the model into the GPU}, while a smaller fraction is spent on model unloading and scheduling. Additionally, the number of model swaps recorded is similar for both settings. However, each swap in CC mode is significantly more time consuming than in No-CC mode, which further contributes to the overall performance gap.

\section{Conclusions and Future Work}
\label{sec:conclusions}

This study investigated the trade-offs between latency, throughput, and GPU utilization in confidential and non-confidential settings. Using state-of-the-art LLMs and hardware, we designed scheduling strategies, generated inference workloads, and systematically measured performance metrics. Our key findings are as follows:
\begin{itemize}
    \item Latency is higher and SLA attainment is lower in CC mode, primarily due to longer model loading times.
    \item Overall throughput and GPU utilization are lower in CC mode, despite the number of model switches being the same in both settings. The performance gap is primarily driven by the increased overhead associated with model loading.
\end{itemize}

In addition to the core performance metrics discussed in Section \ref{sec:experiments}, we used a monitoring tool~\cite{Isenko_Basic_Hardware_Monitor_2023} to track a comprehensive set of system metrics, including:

\begin{itemize}
    \item GPU memory usage: Allocation, peak usage, fragmentation ratio, and memory overhead.
    \item GPU compute utilization.
    \item GPU temperature readings.
    \item CPU metrics: Load, interrupts, and context switches.
    \item Disk activity: Read/write operations.
    \item Network bandwidth usage for sent and received data.
\end{itemize}

This dataset offers rich insights into system performance differences between CC and No-CC settings. Future work will leverage these metrics to explore:
\begin{itemize}
    \item Optimized scheduling strategies that minimize model loading overhead in CC environments.
    \item Alternative inference batching techniques to improve latency and throughput.
    \item Performance evaluations with different LLM architectures and hardware configurations.
\end{itemize}

These directions aim to further bridge the performance gap between confidential and non-confidential AI inference environments while maintaining robust security guarantees.

\bibliographystyle{IEEEtran}
\bibliography{IEEEabrv, Bibliography}

% Generated by IEEEtran.bst, version: 1.12 (2007/01/11)
\begin{thebibliography}{10}
\providecommand{\url}[1]{#1}
\csname url@samestyle\endcsname
\providecommand{\newblock}{\relax}
\providecommand{\bibinfo}[2]{#2}
\providecommand{\BIBentrySTDinterwordspacing}{\spaceskip=0pt\relax}
\providecommand{\BIBentryALTinterwordstretchfactor}{4}
\providecommand{\BIBentryALTinterwordspacing}{\spaceskip=\fontdimen2\font plus
\BIBentryALTinterwordstretchfactor\fontdimen3\font minus \fontdimen4\font\relax}
\providecommand{\BIBforeignlanguage}[2]{{%
\expandafter\ifx\csname l@#1\endcsname\relax
\typeout{** WARNING: IEEEtran.bst: No hyphenation pattern has been}%
\typeout{** loaded for the language `#1'. Using the pattern for}%
\typeout{** the default language instead.}%
\else
\language=\csname l@#1\endcsname
\fi
#2}}
\providecommand{\BIBdecl}{\relax}
\BIBdecl

\bibitem{Confidential_Computing_2022a}
\BIBentryALTinterwordspacing
C.~C. Consortium, ``Confidential computing: Hardware-based trusted execution for applications and data,'' Nov. 2022. [Online]. Available: \url{https://confidentialcomputing.io/wp-content/uploads/sites/10/2023/03/CCC_outreach_whitepaper_updated_November_2022.pdf}
\BIBentrySTDinterwordspacing

\bibitem{Confidential_Computing_2022b}
\BIBentryALTinterwordspacing
------, ``A technical analysis of confidential computing,'' Nov. 2022. [Online]. Available: \url{https://confidentialcomputing.io/wp-content/uploads/sites/10/2023/03/CCC-A-Technical-Analysis-of-Confidential-Computing-v1.3_unlocked.pdf}
\BIBentrySTDinterwordspacing

\bibitem{Confidential_Computing_2022c}
\BIBentryALTinterwordspacing
------, ``Common terminology for confidential computing,'' Dec. 2022. [Online]. Available: \url{https://confidentialcomputing.io/wp-content/uploads/sites/10/2023/03/Common-Terminology-for-Confidential-Computing.pdf}
\BIBentrySTDinterwordspacing

\bibitem{Linux_Foundation_2024}
\BIBentryALTinterwordspacing
T.~L. Foundation, ``The case for confidential computing: Delivering business value through protected, confidential data processing,'' Jul. 2024. [Online]. Available: \url{https://www.linuxfoundation.org/hubfs/LF%20Research/TheCaseforConfidentialComputing_062724.pdf?hsLang=en}
\BIBentrySTDinterwordspacing

\bibitem{Naveed_Khan_Qiu_Saqib_Anwar_Usman_Akhtar_Barnes_Mian_2024}
\BIBentryALTinterwordspacing
H.~Naveed \emph{et~al.}, ``A comprehensive overview of large language models,'' \emph{arXiv}, no. arXiv:2307.06435, Oct. 2024, unpublished. [Online]. Available: \url{http://arxiv.org/abs/2307.06435}
\BIBentrySTDinterwordspacing

\bibitem{IntelTdx}
\BIBentryALTinterwordspacing
P.~Cheng \emph{et~al.}, ``Intel {TDX} demystified: {A} top-down approach,'' \emph{{ACM} Comput. Surv.}, vol.~56, no.~9, pp. 238:1--238:33, 2024. [Online]. Available: \url{https://doi.org/10.1145/3652597}
\BIBentrySTDinterwordspacing

\bibitem{NVIDIA_H100_2023}
\BIBentryALTinterwordspacing
NVIDIA, ``Nvidia h100 tensor core gpu architecture,'' 2023. [Online]. Available: \url{https://resources.nvidia.com/en-us-tensor-core?ncid=no-ncid}
\BIBentrySTDinterwordspacing

\bibitem{secureboot}
\BIBentryALTinterwordspacing
W.~A. Arbaugh, D.~J. Farber, and J.~M. Smith, ``A secure and reliable bootstrap architecture,'' in \emph{1997 {IEEE} Symposium on Security and Privacy, May 4-7, 1997, Oakland, CA, {USA}}.\hskip 1em plus 0.5em minus 0.4em\relax {IEEE} Computer Society, 1997, pp. 65--71. [Online]. Available: \url{https://doi.org/10.1109/SECPRI.1997.601317}
\BIBentrySTDinterwordspacing

\bibitem{remoteattestation}
\BIBentryALTinterwordspacing
G.~Coker \emph{et~al.}, ``Principles of remote attestation,'' \emph{Int. J. Inf. Sec.}, vol.~10, no.~2, pp. 63--81, 2011. [Online]. Available: \url{https://doi.org/10.1007/s10207-011-0124-7}
\BIBentrySTDinterwordspacing

\bibitem{attestation2}
V.~Haldar, D.~Chandra, and M.~Franz, ``Semantic remote attestation: A virtual machine directed approach to trusted computing,'' in \emph{USENIX Virtual Machine Research and Technology Symposium}, vol. 2004.\hskip 1em plus 0.5em minus 0.4em\relax USENIX Association, 2004.

\bibitem{NVIDIA_Hopper_2023}
\BIBentryALTinterwordspacing
NVIDIA, ``Nvidia hopper gpu architecture,'' 2023. [Online]. Available: \url{https://www.nvidia.com/en-us/data-center/technologies/hopper-architecture/}
\BIBentrySTDinterwordspacing

\bibitem{Mohan_Ye_Franke_Srivatsa_Liu_Gonzalez_2024}
\BIBentryALTinterwordspacing
A.~Mohan, M.~Ye, H.~Franke, M.~Srivatsa, Z.~Liu, and N.~M. Gonzalez, ``Securing ai inference in the cloud: Is cpu-gpu confidential computing ready?'' in \emph{2024 IEEE 17th International Conference on Cloud Computing (CLOUD)}, Jul. 2024, pp. 164--175. [Online]. Available: \url{https://ieeexplore.ieee.org/document/10643934/?arnumber=10643934}
\BIBentrySTDinterwordspacing

\bibitem{Mo_Tarkhani_Haddadi_2024}
F.~Mo, Z.~Tarkhani, and H.~Haddadi, ``Machine learning with confidential computing: A systematization of knowledge,'' \emph{ACM Comput. Surv.}, vol.~56, no.~11, pp. 281:1--281:40, Jun. 2024.

\bibitem{Akram_Akella_Peisert_Lowe-Power_2022}
\BIBentryALTinterwordspacing
A.~Akram, V.~Akella, S.~Peisert, and J.~Lowe-Power, ``Sok: Limitations of confidential computing via tees for high-performance compute systems,'' in \emph{2022 IEEE International Symposium on Secure and Private Execution Environment Design (SEED)}, Sep. 2022, pp. 121--132. [Online]. Available: \url{https://ieeexplore.ieee.org/document/9935045/?arnumber=9935045}
\BIBentrySTDinterwordspacing

\bibitem{Zhu_Yin_Deng_Zhou_2024}
\BIBentryALTinterwordspacing
J.~Zhu, H.~Yin, P.~Deng, and S.~Zhou, ``Confidential computing on nvidia h100 gpu: A performance benchmark study,'' \emph{arXiv}, no. arXiv:2409.03992, Sep. 2024, unpublished. [Online]. Available: \url{http://arxiv.org/abs/2409.03992}
\BIBentrySTDinterwordspacing

\bibitem{Li_Zheng_Zhong_Liu_Sheng_Jin_Huang_Chen_Zhang_Gonzalez_2023}
\BIBentryALTinterwordspacing
Z.~Li \emph{et~al.}, ``{{AlpaServe}}: {{Statistical Multiplexing}} with {{Model Parallelism}} for {{Deep Learning Serving}},'' in \emph{17th USENIX Symposium on Operating Systems Design and Implementation}, 2023, pp. 663--679. [Online]. Available: \url{https://www.usenix.org/conference/osdi23/presentation/li-zhouhan}
\BIBentrySTDinterwordspacing

\bibitem{Wang_Wang_Gao_Zhang_Chen_Ng_Ooi_2018}
W.~Wang \emph{et~al.}, ``Rafiki: Machine learning as an analytics service system,'' \emph{Proceedings of the VLDB Endowment}, vol.~12, 04 2018.

\bibitem{NSDI17}
\BIBentryALTinterwordspacing
D.~Crankshaw, X.~Wang, G.~Zhou, M.~J. Franklin, J.~E. Gonzalez, and I.~Stoica, \emph{Proceedings of the 14th {{USENIX Symposium}} on {{Networked Systems Design}} and {{Implementation}} ({{NSDI}} '17): {{Boston}}, {{MA}}, {{USA}}, {{March}} 27 - 29, 2017}.\hskip 1em plus 0.5em minus 0.4em\relax USENIX Association, 2017. [Online]. Available: \url{https://www.usenix.org/system/files/conference/nsdi17/nsdi17-crankshaw.pdf}
\BIBentrySTDinterwordspacing

\bibitem{Ye_Gao_Hu_Sun_Wang_Luo_Zhang_Wen_2024}
Z.~Ye \emph{et~al.}, ``Deep learning workload scheduling in gpu datacenters: A survey,'' \emph{ACM Computing Surveys}, vol.~56, no.~6, pp. 1--38, Jun. 2024.

\bibitem{instructlab}
\BIBentryALTinterwordspacing
 [Online]. Available: \url{https://github.com/instructlab}
\BIBentrySTDinterwordspacing

\bibitem{HuggingFaceAI2024}
\BIBentryALTinterwordspacing
Hugging {{Face}} – {{The AI}} community building the future. [Online]. Available: \url{https://huggingface.co/}
\BIBentrySTDinterwordspacing

\bibitem{Isenko_Basic_Hardware_Monitor_2023}
\BIBentryALTinterwordspacing
A.~Isenko. (2023, Sep.) {Basic Hardware Monitor}. [Online]. Available: \url{https://github.com/cirquit/py-hardware-monitor}
\BIBentrySTDinterwordspacing

\end{thebibliography}

\end{document}